\documentclass[aps,preprint,showpacs,floats,epsf,epsfig,nofootinbib,12pt]{revtex4}
\usepackage{graphicx}% Include figure files
\usepackage{epsfig}
\usepackage[dvips]{color}

\def\beq{\begin{eqnarray}}
\def\eeq{\end{eqnarray}}
\def\non{\nonumber}

\def\la{\langle}
\def\ra{\rangle}
\def\Mb{M_{\Sigma_b}}
\def\Mc{M_{\Sigma_c}}

\begin{document}

\title{  $\Sigma_{b}\to\Sigma_c$ and $\Omega_b\to\Omega_c$ weak decays in the light-front quark model }

\vspace{1cm}

\author{ Hong-Wei Ke$^{1}$   \footnote{khw020056@hotmail.com},
          Xu-Hao Yuan$^{2}$ \footnote{yuanxh@tsinghua.edu.cn},
        Xue-Qian Li$^3$\footnote{lixq@nankai.edu.cn},
        Zheng-Tao Wei$^3$\footnote{weizt@nankai.edu.cn} and
        Yan-Xi Zhang $^{2}$ \footnote{yanxi.zhang@cern.ch}}

\affiliation{  $^{1}$ School of Science, Tianjin University, Tianjin 300072, China
\\ $^{2}$ Center for High Energy Physics, Department of Engineering Physics,
Tsinghua University, Beijing 100084, China
\\
  $^{3}$ School of Physics, Nankai University, Tianjin 300071, China }

\vspace{12cm}

\begin{abstract}

The successful operation of LHC provides a great opportunity to
study the processes where heavy baryons are involved. {In this
work we mainly study} the weak transitions of $\Sigma_b\to
\Sigma_c$. Assuming the reasonable quark-diquark  structure where
the two light quarks constitute an axial vector, we calculate the
widths of semi-leptonic decay $\Sigma_{b}\to\Sigma_c e\nu_e$ and
non-leptonic decay modes $\Sigma_{b}\to\Sigma_c +M$ (light mesons)
in terms of the light front quark model. We first construct the
vertex function for the concerned baryons and then deduce the form
factors which  are related to two Isgur-Wise functions for the
$\Sigma_{b}\to\Sigma_c$ transition under the heavy quark limit.
Our numerical results indicate that $\Gamma(\Sigma_{b}\to\Sigma_c
e\nu_e)$ is about $1.38\times10^{10}{\rm s}^{-1}$ and
$\Gamma(\Sigma_{b}\to\Sigma_c +M)$ is slightly below
$1\times10^{10}{\rm s}^{-1}$ which may be accessed at the LHCb
detector. By the flavor SU(3) symmetry  we estimate the rates of
$\Omega_b\to\Omega_c$. We suggest to measure weak decays of
$\Omega_b\to\Omega_c$, because $\Omega_b$ does not decay via
strong interaction, the advantage is obvious.

\pacs{13.30.-a, 12.39.Ki, 14.20.Lq, 14.20.Mr}

\end{abstract}

\maketitle

\section{Introduction}

In our previous work \cite{Ke:2007tg}, we investigated the
transitions between heavy baryons $\Lambda_b\to\Lambda_c$ by
assuming the baryonic heavy-quark-light-diquark structures in
terms of the Light-Front-Quark model (LFQM). The results are
reasonably consistent with the available data, so it implies that
the whole scenario is realistic in that case, however still needs
more studies on its validity in other cases. As noticed that the
ground state diquarks in $\Lambda_b$ and $\Lambda_c$ are
color-anti-triplet scalars. In this work, we continue to consider
the transitions of $\Sigma_b\to\Sigma_c$ because the ground state
diquark in $\Sigma_{b(c)}$ is an axial vector. We explore if the
difference of the diquark identities would result in distinct
behaviors for the transitions and then by comparing with data we
are able to gain more insight about the diquark structure.

Thanks to the successful operation of LHC, a remarkable database on
baryons, especially on the heavy baryons will be available at LHCb.
It enables researchers to closely study the properties of heavy
baryons at their production and decay processes.

Since the situation is confronting a radical change, more physicists
are turning to concern baryons and look for hints of new physics.
For example, as the decay $\Sigma\rightarrow p\,\mu^+\,\mu^-$ was
observed\cite{Park:2005eka} the authors of
Ref.\cite{He:2006fr,He:2005we} studied contribution from new physics
candidates by analyzing the data. However, as it is well known when
one explores possible new physics scenario based on the data, he
needs to fully understand the contribution of the standard model
(SM) i.e. before attributing the phenomena to new physics  a
complete analysis on the SM contribution is necessary.

In this work we explore the weak transition of
$\Sigma_{b}\rightarrow \Sigma_{c}$. The dominant strong decay mode
$\Sigma_b\rightarrow\Lambda_b+\pi$ determines the lifetime of
$\Sigma_b$, thus the weak decays of $\Sigma_b$ are rare. However,
from another aspect, the rare decays of $\Sigma_b$ may be more
sensitive to new physics, so that it is worth a careful study.
%\textcolor[rgb]{0.00,1.00,0.00}{Moreover, the CKM parameter
%$V_{cb}$ can extract from the weak decay of $\Sigma_{b}\rightarrow
%\Sigma_{c}$.}
%and especially could serve as an ideal
%laboratory to investigate the non-perturbative QCD effects for the
%baryon case.

Supposing the factorization is valid, the transition between quarks
would be fully described by the perturbative theory and calculable,
thus the main task for studying $\Sigma_{b}\rightarrow \Sigma_{c}$
is to deal with the hadronic transition matrix element. The hadronic
matrix elements are determined by non-perturbative QCD and are
generally parameterized by some form factors which can be reduced
into a few equivalent Isgur-Wise functions under the heavy quark
limit\cite{IW}. Some authors
\cite{Korner:1992uw,Ebert:2006rp,Singleton:1990ye,Ivanov:1996fj,Ivanov:1998ya}
calculated the form factors of the transition $\Sigma_{b}\rightarrow
\Sigma_{c}$ in various approaches.

The quark-diquark structure that heavy baryons are made of a heavy
quark and a light diquark\cite{wilczek,GKLLW,yu} is generally
considered as a reasonable physics picture for heavy baryons. With
the quark-diquark structure the authors of
Refs.\cite{Korner:1992uw,Ebert:2006rp,Ke:2007tg,Wei:2009np}
evaluated the transition rates between heavy baryons   and their
results are consistent with the available data. It is noted that
the diquark stands as a spectator in the transition of
$\Sigma_{b}\rightarrow \Sigma_{c}$, so that under the heavy quark
limit, the spin of light diquark decouples and we may evaluate the
rates of the corresponding rare decays in terms of the Isgur-Wise
functions. A general analysis suggests that there exist many
Isgur-Wise-type functions for a transition between baryons
\cite{Guo} and usually it would be hard to determine them by
fitting data. However, with the quark-diquark structure, the
number of such functions for the transition $\Sigma_b\to\Sigma_c$
reduces into only two.

The light-front quark model (LFQM) is a relativistic quark model
which has been applied to study transitions among mesons and the
results agree with the data within reasonable error tolerance
\cite{Jaus,Ji:1992yf,Cheng:1996if,Cheng:2003sm,Hwang:2006cua,Li:2010bb,Ke:2009ed,Wei:2009nc,Choi:2007se},
thus we would be tempted to extend its application to calculate the
transition of $\Sigma_{b}\rightarrow \Sigma_{c}$ as long as the
diquark picture is employed. In Ref.\cite{Ke:2007tg} we calculated
the transition of $\Lambda_b\rightarrow\Lambda_c$ in terms of LFQM.
In that work we first constructed the vertex function of
$\Lambda_{b(c)}$ and then deduced the form factors for the
transition. However the formulas in \cite{Ke:2007tg} do not apply to
the decay $\Sigma_{b}\rightarrow \Sigma_{c}$  because the diquark in
$\Lambda_{b(c)}$ is a scalar of color-anti-triplet, but that in
$\Sigma_{b(c)}$ is an axial vector as discussed in
Refs.\cite{Korner:1992uw,Ebert:2006rp}. Thus we need to re-construct
the vertex function for a $\frac{1}{2}^+$ heavy baryon which is
regarded as a bound state of a heavy quark and a light axial vector
diquark. Then with the vertex functions of baryons we would derive
the transition matrix element which are parametrized by a few form
factors, and  under the heavy quark limit, we will show that the
transition matrix element of $\Sigma_b\to\Sigma_c$ can be described
by two generalized Isgur-Wise functions. Numerically the results
obtained in the two approaches are rather close, so it implies that
the employed approaches are reasonably consistent with the physical
picture.

Since the leptons do not participate in the strong interaction,
the semileptonic decay is simple and less contaminated by the
non-perturbative QCD effect, therefore study on semileptonic decay
might help to test the employed model and/or constrain the model
parameters. With the form factors we evaluate the width of the
semileptonic decay. Comparing our numerical result with data the
model parameters which are hidden in the vertex functions can be
fixed. Moreover, the amplitude of the non-leptonic decay
$\Sigma_{b}\rightarrow \Sigma_{c}+M$ can also be evaluated in a
similar way as long as we suppose that the meson current can be
factorized out. Moreover, we further investigate the transitions
of $\Omega_b\to \Omega_c$ by assuming the flavor SU(3) symmetry.
Since $\Omega_b$ does not decay via strong interaction, the weak
decays are dominant, so that study on such modes has an obvious
advantage.

This paper is organized as follows: after the introduction, in
section II we construct the vertex functions of heavy baryons, then
derive the form factors for the transition $\Sigma_{b}\rightarrow
\Sigma_{c}$ in the light-front quark model, then we present our
numerical results for the transition $\Sigma_{b}\rightarrow
\Sigma_{c}$ along with all necessary input parameters in section
III, then we also evaluate the transition of $\Omega_b\to \Omega_c$.
Section IV is devoted to our conclusion and discussions.

\section{$\Sigma_{b}\rightarrow \Sigma_{c}$ in the light-front quark model}

By  the quark-diquark structure\cite{Korner:1992uw,Ebert:2006rp},
the heavy baryon $\Sigma_{b(c)}$ consists of a light $1^+$ diquark
[ud] and one heavy quark $b(c)$.
%In order to form a color singlet hadron, the
%diquark [ud] is a color anti-triplet state.
To insure the quantum number of $\Sigma_{b(c)}$, the orbital angular
momentum between the two components is zero, i.e. $l=0$.

\subsection{the vertex function of $\Sigma_{b(c)}$ }

In analog to our previous work \cite{Ke:2007tg}, we construct the
vertex function of $\Sigma_Q$ ($Q=b, c$) where the diquark is an
axial vector in the same model.  The wavefunction of $\Sigma_Q$ with
total spin $S=1/2$ and momentum $P$ is
\begin{eqnarray}\label{eq:lfbaryon}
 |\Sigma_Q(P,S,S_z)\rangle&=&\int\{d^3\tilde p_1\}\{d^3\tilde p_2\} \,
  2(2\pi)^3\delta^3(\tilde{P}-\tilde{p_1}-\tilde{p_2}) \nonumber\\
 &&\times\sum_{\lambda_1}\Psi^{SS_z}(\tilde{p}_1,\tilde{p}_2,\lambda_1)
  C^{\alpha}_{\beta\gamma}F^{bc}\left|\right.
  Q_{\alpha}(p_1,\lambda_1)[q_{1b}^{\beta}q_{2c}^{\gamma}](p_2)\ra,
\end{eqnarray}
with
\begin{eqnarray*}
 \Psi^{SS_z}(\tilde{p}_1,\tilde{p}_2,\lambda_1,\lambda_2)=&&
  \left\la\lambda_1\left|\mathcal{R}^{\dagger}_M(x_1,k_{1\perp},m_1)
   \right|s_1\right\ra  \left\la\lambda_2\left|\mathcal{R}^{\dagger}_M(x_2,k_{2\perp},m_2)
   \right|s_2\right\ra
  \nonumber\\ &&\left\la
\frac{1}{2}s_1 ;1 s_2\left|\frac{1}{2}S_z\right\ra
 \varphi(x,k_{\perp})\right.,
\end{eqnarray*}
where $ \langle\frac{1}{2}s_1;1  s_2|\frac{1}{2}S_z\rangle$ is the
C-G coefficients and $s_1,s_2$ are the spin projections of the
constituents (the heavy quark and diquark).  A Melosh transformation
brings the the matrix elements from the
spin-projection-on-fixed-axes representation into the helicity
representation and  is explicitly written as
$$\left\la\lambda_2\left|\mathcal{R}^{\dagger}_M(x_2,k_{2\perp},m_2)
   \right|s_2\right\ra=\xi^*(\lambda_1,m_2)\cdot\xi(s_2,m_2),$$ and $$\left\la\lambda_1\left|\mathcal{R}^{\dagger}_M(x_1,k_{1\perp},m_1)
   \right|s_1\right\ra=\frac{\bar{u}(k_1,\lambda_1)u(k_1,s_1)}{2m_1}.$$

Following Refs. \cite{Jaus,Cheng:2003sm}, the Melosh transformed
matrix can be expressed as
\begin{eqnarray}
 &&\left\la\lambda_2\left|\mathcal{R}^{\dagger}_M(x_2,k_{2\perp},m_2)
   \right|s_2\right\ra\left\la\lambda_1\left|\mathcal{R}^{\dagger}_M(x_1,k_{1\perp},m_1)
   \right|s_1\right\ra \left\la \frac{1}{2}s_1;1 s_2\left|
   \frac{1}{2}S_z\right\ra\right.\nonumber\\&&=\frac{1}{\sqrt{2(p_1\cdot
   \bar P+m_1M_0)}}\bar{u}(p_1,\lambda_1)\Gamma u(\bar {P},S_z),
\end{eqnarray}
where
\begin{eqnarray}
 \Gamma=-\frac{1}{\sqrt{3}}\gamma_5
\varepsilon^* \!\!\!\!\!\!\slash \; (p_2,\lambda_2),\qquad
m_1=m_{Q},\qquad m_2=m_{[ud]}, \qquad \bar P=p_1+p_2,
\end{eqnarray}
and
\begin{eqnarray}
&&\varphi(x,k_\perp)=A\phi,
\end{eqnarray}
with
$\phi=4(\frac{\pi}{\beta^2})^{3/4}\frac{e_1e_2}{x_1x_2M_0}{\rm
exp}(\frac{-\mathbf{k}^2}{2\beta^2})$ and
$A=\sqrt{\frac{12(M_0m_1+p_1\cdot\bar{P})}{12M_0m_1+4p_1\bar{P}+8p_1\cdot
p_2 p_2\cdot \bar{P}/m_2^2}}$ which can be obtained by normalizing
the state  $|\Sigma_Q(P,S,S_z)\rangle$ , \beq\label{A12}
 \la
 \Sigma_Q(P',S',S'_z)|\Sigma_Q(P,S,S_z)\ra=2(2\pi)^3P^+
  \delta^3(\tilde{P}'-\tilde{P})\delta_{S'S}\delta_{S'_zS_z}.
 \eeq

{ All other notations can be found in Ref}.\cite{Ke:2007tg}.
\subsection{$\Sigma_b\to\Sigma_c$ transition form factors }

\begin{figure}
\begin{center}
%\begin{tabular}{ccc}
\scalebox{0.8}{\includegraphics{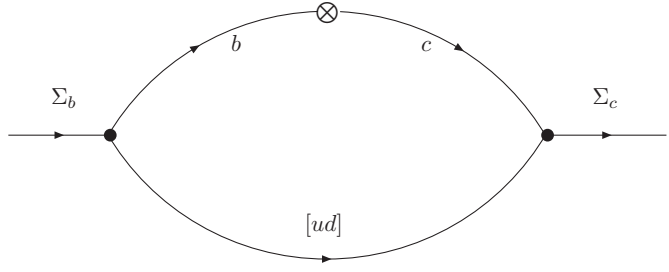}}
%\end{tabular}
\end{center}
\caption{Feynman diagram for $\Sigma_{b}\to\Sigma_{c}$
transitions, where $\bigotimes$ denotes $V-A$ current
vertex.}\label{t1}
\end{figure}

The lowest order Feynman diagram for the $\Sigma_{b}\to\Sigma_{c}$
weak decay is shown in Fig. \ref{t1}. %In Ref.\cite{pentaquark2}, the
%light-front quark model for heavy pentaquark with one heavy quark
%and two light diquarks is presented.

%Thus, most of our formulations are similar to
%that in \cite{pentaquark2} except there is only one diquark in our
%case.
Using  the wavefunction for $\mid \Sigma_{Q}(P,S,S_z) \ra$, we
obtain
\begin{eqnarray}\label{s2}
&& \la \Sigma_{Q'}(P',S_z') \mid \bar{Q}'
\gamma^{\mu} (1-\gamma_{5}) Q \mid \Sigma_{Q}(P,S_z) \ra  \nonumber \\
 &=& \int\{d^3 \tilde p_2\}\frac{\phi_{\Sigma_{Q'}}^*(x',k'_{\perp})
  \phi_{\Sigma_Q}(x,k_{\perp})}{2\sqrt{p^+_1p'^+_1(p_1\cdot \bar{P}+m_1M_0)
 (p'_1\cdot \bar{P'}+m'_1M'_0)}}\nonumber \\
  &&\times \bar{u}(\bar{P'},S'_z)\bar{\Gamma}'
  (p_1\!\!\!\!\!\slash'+m'_1)\gamma^{\mu}(1-\gamma_{5})
  (p_1\!\!\!\!\!\slash+m_1)\Gamma u(\bar{P},S_z),
\end{eqnarray}
where
 \beq
 &&\bar{\Gamma}'=\gamma_0\Gamma'\gamma_0, \non \\
 &&m_1=m_b, \qquad m'_1=m_c, \qquad m_2=m_{[ud]},
 \eeq
{and $Q$($Q'$) represent $b$($c$) quark, $p_1$($p'_1$) is its
momentum and  $P$($P'$) denotes the momentum of initial (final)
baryon. } From $\tilde{p}_2=\tilde{p}'_2$, we have
 \beq
 x'=\frac{P^+}{P^{'+}}x, \qquad \qquad
 k'_{\perp}=k_{\perp}+x_2q_{\perp}.
 \eeq
with $x=x_2$, $x'=x'_2$. Thus, Eq. (\ref{s2}) is rewritten as
\begin{eqnarray}\label{s23}
 &&\la \Sigma_{Q'}(P',S_z') \mid \bar{Q}' \gamma^{\mu}
  (1-\gamma_{5}) Q \mid \Sigma_{Q}(P,S_z) \ra \non\\
  &=& \int\frac{dxd^2k_{\perp}}{2(2\pi)^3}\frac{
  \phi_{\Sigma_{Q'}}(x',k'_{\perp})
  \phi_{\Sigma_Q}(x,k_{\perp})}
  {2\sqrt{x_1x'_1(p_1\cdot \bar{P}+m_1M_0)
  (p'_1\cdot \bar{P'}+m'_1M'_0)}}\nonumber \\
 &&\times
 \bar{u}(\bar{P'},S'_z)[\frac{1}{\sqrt{3}}\gamma_5\varepsilon(\lambda_2')^*]
  (p_1\!\!\!\!\!\slash'+m'_1)\gamma^{\mu}(1-\gamma_{5})
  (p_1\!\!\!\!\!\slash+m_1) [\frac{1}{\sqrt{3}}\gamma_5\varepsilon(\lambda_2)] u(\bar{P},S_z).
\end{eqnarray}

The form factors for the weak transition $\Sigma_Q\rightarrow
\Sigma_{Q'}$  are defined in the standard way as
\begin{eqnarray}\label{s1}
&& \la \Sigma_{Q'}(P',S',S_z') \mid \bar{Q}'\gamma_{\mu}
 (1-\gamma_{5})Q \mid \Sigma_{Q}(P,S,S_z) \ra  \non \\
 &=& \bar{u}_{\Sigma_{Q'}}(P',S'_z) \left[ \gamma_{\mu} f_{1}(q^{2})
 +i\sigma_{\mu \nu} \frac{ q^{\nu}}{M_{\Sigma_{Q}}}f_{2}(q^{2})
 +\frac{q_{\mu}}{M_{\Sigma_{Q}}} f_{3}(q^{2})
 \right] u_{\Sigma_{Q}}(P,S_z) \nonumber \\
 &&-\bar u_{\Sigma_{Q'}}(P',S'_z)\left[\gamma_{\mu} g_{1}(q^{2})
  +i\sigma_{\mu \nu} \frac{ q^{\nu}}{M_{\Sigma_{Q}}}g_{2}(q^{2})+
  \frac{q_{\mu}}{M_{\Sigma_{Q}}}g_{3}(q^{2})
 \right]\gamma_{5} u_{\Sigma_{Q}}(P,S_z).
\end{eqnarray}
where  $q \equiv P-P'$, $Q$ and $Q'$ denote $b$ and $c$,
respectively.  Since $S=S'=1/2$, we will be able to write
$\mid\Sigma_{Q}(P,S,S'_z)\ra$ as $\mid\Sigma_{Q}(P,S_z)\ra$.

Following\cite{Ke:2007tg,pentaquark2}, we extract the form factors
for the weak transition matrix elements of $\Sigma_b\to \Sigma_c$ as
\begin{eqnarray}\label{s6}
f_1(q^2)&=&\frac{1}{8P^+P'^+}\int{\frac{dx_2d^2k_{2\perp}}{2(2\pi)^3}}
\frac{\phi'^*_{00}(x',k'_{\perp})\phi_{00}(x,k_{\perp})}{6\sqrt{x_1x'_1(p_1\cdot
\bar{P}+m_1M_0)(p'_1\cdot \bar{P'}+m'_1M'_0)}} \nonumber
\\&&Tr[(\bar{P}\!\!\!\!\slash+M_0)\gamma^+(\bar{P'}\!\!\!\!\!\!\slash+M'_0)
\gamma_5\gamma_a(p_1\!\!\!\!\!\slash'+m'_1)\gamma^{+}
(p_1\!\!\!\!\!\slash+m_1)\gamma_5\gamma_b](\frac{p_2^ap_2^b}{m_2^2}-g^{ab})\nonumber \\
g_1(q^2)&=&\frac{1}{8P^+P'^+}\int{\frac{dx_2d^2k_{2\perp}}{2(2\pi)^3}}
\frac{\phi'^*_{00}(x',k'_{\perp})\phi_{00}(x,k_{\perp})}{6\sqrt{x_1x'_1(p_1\cdot
\bar{P}+m_1M_0)(p'_1\cdot \bar{P'}+m'_1M'_0)}} \nonumber
\\&&Tr[(\bar{P}\!\!\!\!\slash+M_0)\gamma^+\gamma_5(\bar{P'}\!\!\!\!\!\!\slash+M'_0)
\gamma_5\gamma_a(p_1\!\!\!\!\!\slash'+m'_1)\gamma^{+}\gamma_5
(p_1\!\!\!\!\!\slash+m_1)\gamma_5\gamma_b](\frac{p_2^ap_2^b}{m_2^2}-g^{ab})\nonumber \\
\frac{f_2(q^2)}{M_{\Lambda_Q}}&=&-\frac{1}{8P^+P'^+q^i_{\perp}}\int{\frac{dx_2d^2k_{2\perp}}{2(2\pi)^3}}
\frac{\phi'^*_{00}(x',k'_{\perp})\phi_{00}(x,k_{\perp})}{6\sqrt{x_1x'_1(p_1\cdot
\bar{P}+m_1M_0)(p'_1\cdot \bar{P'}+m'_1M'_0)}} \nonumber
\\&&Tr[(\bar{P}\!\!\!\!\slash+M_0)\sigma^{i+}(\bar{P'}\!\!\!\!\!\!\slash+M'_0)
\gamma_5\gamma_a(p_1\!\!\!\!\!\slash'+m'_1)\gamma^{+}
(p_1\!\!\!\!\!\slash+m_1)\gamma_5\gamma_b](\frac{p_2^ap_2^b}{m_2^2}-g^{ab})\nonumber \\
\frac{g_2(q^2)}{M_{\Lambda_Q}}&=&\frac{1}{8P^+P'^+q^i_{\perp}}\int{\frac{dx_2d^2k_{2\perp}}{2(2\pi)^3}}
\frac{\phi'^*_{00}(x',k'_{\perp})\phi_{00}(x,k_{\perp})}{6\sqrt{x_1x'_1(p_1\cdot
\bar{P}+m_1M_0)(p'_1\cdot \bar{P'}+m'_1M'_0)}} \nonumber
\\&&Tr[(\bar{P}\!\!\!\!\slash+M_0)\sigma^{i+}\gamma_5(\bar{P'}\!\!\!\!\!\!\slash+M'_0)
\gamma_5\gamma_a(p_1\!\!\!\!\!\slash'+m'_1)\gamma^{+}\gamma_5
(p_1\!\!\!\!\!\slash+m_1)\gamma_5\gamma_b](\frac{p_2^ap_2^b}{m_2^2}-g^{ab}).
\end{eqnarray}
with $i=1,2$. The traces can be worked out straightforwardly and
all the details can be found in Ref.\cite{Ke:2007tg}.

\subsection{ Isgur-Wise functions of the transition}

As  well known  under the heavy quark limit
($m_Q\to\infty$)\cite{HQS},  the six form factors $f_i,~g_i$
(i=1,2,3) are no longer independent, but are related to each other
by an extra symmetry. Thus the matrix elements are determined by two
universal Isgur-Wise functions $\xi_1(v\cdot v')$ and $\xi_2(v\cdot
v')$.

The generalized Isgur-Wise functions in the
$\Sigma_Q\to\Sigma_{Q'}$ transition are defined through the
following expression
  \begin{eqnarray}\label{s10}
  && < \Sigma_{Q'}(v',S_z')\mid\bar{Q}_{v'}' \gamma_{\mu}
   (1-\gamma_{5}) Q_v \mid \Sigma_{Q}(v,S_z)>\nonumber\\&& = \frac{1}{3} [g^{\alpha\beta}\xi_1(\omega)-v^\alpha v'^\beta\xi_2(\omega)]
  \bar{u}(v',S'_z)\gamma_5(\gamma_\alpha+v'_\alpha)\gamma_{\mu}(1-\gamma_{5})(\gamma_\beta+v_\beta)\gamma_5
  u(v,S_z),
  \end{eqnarray}
where $\omega\equiv v\cdot v'$. In fact, as we re-calculate the
transition matrix elements under the heavy quark limit, one can
easily obtain a new expression corresponding to Eq.(9) where there
are six independent form factors.

%    \begin{eqnarray}
%  &&f_1=F_1+\frac{1}{2}(\frac{M_Q+M_Q'}{M_Q}F_2+\frac{M_Q+M_Q'}{M_Q'}F_3),
%   g_1=G_1-\frac{1}{2}(\frac{M_Q-M_Q'}{M_Q}G_2+\frac{M_Q-M_Q'}{M_Q'}G_3)\nonumber\\&&
 %  f_2=\frac{1}{2}(F_2+\frac{M_Q}{M_Q'}F_3), g_2=\frac{1}{2}(G_2+\frac{M_Q}{M_Q'}G_3)\nonumber\\&&
%    f_3=\frac{1}{2}(F_2-\frac{M_Q}{M_Q'}F_3), g_3=\frac{1}{2}(G_2-\frac{M_Q}{M_Q'}G_3)
 %     \end{eqnarray}
%   with
 %     \begin{eqnarray}
 %   &&F_1=G_1=-\frac{1}{3}[\omega \xi_1+(1-\omega^2)\xi_2]\nonumber\\&&
 %    F_2=G_2=\frac{2}{3}[\omega \xi_1+(1-\omega)\xi_2]\nonumber\\&&
 %     F_3=G_3=\frac{2}{3}[\omega \xi_1-(1+\omega)\xi_2]
 %       \end{eqnarray}

As discussed in Ref.\cite{Ke:2007tg}  with a replacements in the
heavy quark effective theory (HQET)
 \beq
 \mid\Sigma_{Q}(P,S_z)\ra &\to& \sqrt{M_{\Sigma_Q}}
 \mid\Sigma_{Q}(v,S_z)\ra,
  \non \\
 u(\bar{P},S_z) &\to& \sqrt{m_Q}u(v,S_z) \non \\
 \phi_{\Sigma_Q}(x,k_{\bot})&\to &\sqrt{\frac{m_Q}{X}}\Phi(X,k_{\bot}),
 \eeq
and
\begin{eqnarray}\label{s12}
 &&M_{\Sigma_Q}\to m_Q, \qquad ~~ M_0\to m_Q,\non\\
 &&e_1\to m_Q, \qquad\qquad \non\\
 &&e_2\to v\cdot p_2=\frac{m_2^2+k_{\perp}^2+X^2}{2X},\non\\
 &&\vec k^2\to (v\cdot p_2)^2-m_2^2,\non\\
 &&p_1\!\!\!\!\!\slash+m_1\to m_Q(v\!\!\!\slash+1)\non \\
 &&\frac{e_1e_2}{x_1x_2M_0}\to\frac{m_Q}{X}(v\cdot p_2),
\end{eqnarray}
we are able to re-formulate the transition form factors obtained in
the previous section under the heavy quark limit.

The  matrix element of the transition $\Sigma_Q\to\Sigma_{Q'}$ is
then
\begin{eqnarray}\label{sw14}
 &&\la\Sigma_{Q'}(v',S_z') \mid \bar{Q}_{v'}'
   \gamma^{\mu}(1-\gamma_{5})Q_v\mid\Sigma_{Q}(v,S_z) \ra  \non\\
 &=& -\int{\frac{dX}{X}\frac{d^2k_{\perp}}{2(2\pi)^3}}
     {\Phi(X,k_{\perp})\Phi(X',k_{\perp}^{\prime})}
    \frac{1}{3} \bar{u}(v',S'_z)\gamma_5(\gamma_\alpha+v'_\alpha)\gamma^\mu(1-\gamma_5)\nonumber\\&&
    (\gamma_\beta+v'_\beta)\gamma_5u(v,S_z)(\frac{p_2^\alpha p_2^\beta}{m_2^2}-g^{\alpha\beta}),
\end{eqnarray}
with
 \beq
 \Phi(X,k_{\bot})=\sqrt{\frac{24}{16+8v\cdot
p_2^2/m_2^2}}4\sqrt{v\cdot p_2}\left(\frac{\pi}{\beta^2_\infty}
  \right)^{\frac{3}{4}}{\rm exp}\left(-\frac{(v\cdot p_2)^2-m^2_2}
  {2\beta^2_\infty}\right),
 \eeq
where $\beta_{\infty}$ denotes the value of $\beta$ in the heavy
quark limit.

Thus we can write down the transition matrix element as
          \begin{eqnarray}\label{s14}
        &&<\Sigma_{Q'}(v',S_z') \mid \bar{Q}_{v'}'
          \gamma^{\mu}(1-\gamma_{5})Q_v\mid\Sigma_{Q}(v,S_z) >
       \nonumber\\&& =- \int{\frac{dX}{X}\frac{d^2k_{\perp}}{2(2\pi)^3}}
        {\Phi(X,k_{\perp})\Phi(X',k_{\perp}^{\prime})} \frac{1}{3} \bar{u}(v',S'_z)
        \gamma_5(\gamma_\alpha+v'_\alpha)\gamma^\mu(1-\gamma_5) \nonumber\\&&
       (\gamma_\beta+v'_\beta)\gamma_5u(v,S_z)(
       a_1 g^{\alpha\beta}+a_2 v^\alpha v'^\beta+a_3 v'^\alpha v^\beta+a_4 v'^\alpha v'^\beta+a_5 v^\alpha v^\beta).
       \end{eqnarray}
By the relation
$\bar{u}'\gamma_5(v'\!\!\!\!\slash+1)=(v\!\!\!\slash+1)\gamma_5u=0$,
the terms with $a_3$, $a_4$ and $a_5$ do not contribute to the
transition, thus
 \begin{eqnarray}\label{s14}
        &&<\Sigma_{Q'}(v',S_z') \mid \bar{Q}_{v'}'
          \gamma^{\mu}(1-\gamma_{5})Q_v\mid\Sigma_{Q}(v,S_z) >
       \nonumber\\&& =- \int{\frac{dX}{X}\frac{d^2k_{\perp}}{2(2\pi)^3}}
        {\Phi(X,k_{\perp})\Phi(X',k_{\perp}^{\prime})} \frac{1}{3} \bar{u}(v',S'_z)
        \gamma_5(\gamma_\alpha+v'_\alpha)\gamma^\mu(1-\gamma_5) \nonumber\\&&
       (\gamma_\beta+v'_\beta)\gamma_5u(v,S_z)(
       a_1 g^{\alpha\beta}+a_2 v^\alpha v'^\beta),
       \end{eqnarray}
and
       \begin{eqnarray}
          && a_1=-\frac{(w^2-1)p_2^2+2v\cdot p_2 v'\cdot p_2 \omega-(v'\cdot p_2)^2
           -(v\cdot p_2)^2 }{2m_2^2(\omega^2-1)} \nonumber\\&&
           a_2=-\frac{\omega(\omega^2-1)p_2^2-2v\cdot p_2 v'\cdot p_2(2\omega^2+1)
           +3\omega[(v'\cdot p_2)^2+(v\cdot p_2)^2]}{2m_2^2(\omega^2-1)^2}
       \end{eqnarray}

Comparing Eq.(\ref{s14}) with Eq. (\ref{s10}), we get
       \begin{eqnarray}\label{s14}
       && \xi_1 = -\int{\frac{dX}{X}\frac{d^2k_{\perp}}{2(2\pi)^3}}
        {\Phi(X,k_{\perp})\Phi(X',k_{\perp}^{\prime})}a_1,
       \end{eqnarray}
        \begin{eqnarray}\label{s14}
        && \xi_2 = \int{\frac{dX}{X}\frac{d^2k_{\perp}}{2(2\pi)^3}}
         {\Phi(X,k_{\perp})\Phi(X',k_{\perp}^{\prime})}a_2.
        \end{eqnarray}
The forms of $\xi_1$ and $\xi_2$ are similar to that in Eq.(4.18)
and Eq. (4.19) of Ref.\cite{pentaquark2} and  can be directly
evaluated in the time-like region by choosing a reference frame
where $q_\perp=0$.

\section{Numerical Results}

In this section we present our numerical results  for the
transition $\Sigma_b\to\Sigma_c$ along with all input parameters.
First we need to obtain the form factors, then using them the
predictions on semi-leptonic processes $\Sigma_b\rightarrow
\Sigma_c l\bar{\nu}_l$ and non-leptonic decays $\Sigma_b\to
\Sigma_c M^-$ ($M$ represents $\pi,~ K,~ \rho,~ K^*,~ a_1$ etc.)
will be made.

First of all, let us list our input parameters. The baryon masses
$\Mb=5.807$ GeV, $\Mc=2.452$ GeV are taken from\cite{PDG10}.  For
the heavy quark masses, we set $m_b$ and $m_c$ following
Ref.\cite{Cheng:2003sm}. In the early literature, the mass of the
constituent light axial vector diquark $m_{[ud]}$ disperses in a
rather wide range, for example, it is set as: 614-618
MeV\cite{Ram:1986fa}, 770 MeV\cite{Korner:1992uw} , 909
MeV\cite{Ebert:2006rp}. In \cite{Ke:2007tg} we fixed the scalar
diquark mass as $m_{[ud]_{_S}}=500$ MeV. Generally an axial vector
should be slightly heavier than a scalar with the same constituents,
so we set $m_{[ud]_{_V}}=770$ MeV. Since the [ud] diquark mass is
close to the mass of a strange quark, we may assume that the
parameters $\beta_{b[ud]}$ and $\beta_{c[ud]}$ should be close to
$\beta_{b\bar s}$ and $\beta_{c\bar s}$ which appear in the meson
case\cite{Cheng:2003sm}. All the input parameters are collected in
Table \ref{Tab:t1}.

\begin{table}
\caption{Quark mass and the parameter $\beta$ (in units of
 GeV).}\label{Tab:t1}
\begin{ruledtabular}
\begin{tabular}{ccccc}
  $m_c$  & $m_b$  &$m_{[ud]}$ & $\beta_{c[ud]}$ & $\beta_{b[ud]}$ \\
  $1.3$  & $4.4$  & 0.77     & $0.45$         & 0.50
\end{tabular}
\end{ruledtabular}
\end{table}

\subsection{$\Sigma_b\to \Sigma_c$ form factors and the Isgur-Wise
functions}

\begin{figure}
\begin{center}
\scalebox{1.0}{\includegraphics{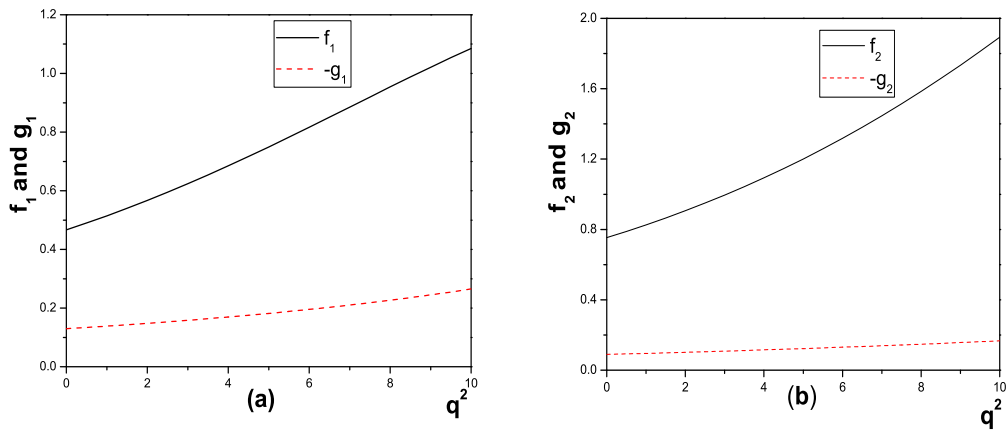}}
\end{center}
\caption{(a) Form factors  $ f_1$ and $g_1$ \,\,\,  (b)
 Form factors  $f_2$ and $g_2$}\label{f2}
\end{figure}

As discussed in Ref.\cite{Cheng:2003sm} the  form factors are
calculated in the frame $q^+=0$ with $q^2=-q^2_{\perp}\leq 0$ (the
space-like region). To extended them into the time-like region, an
analytic three-parameter form was suggested \cite{pentaquark2}
 \begin{eqnarray}\label{s14}
 F(q^2)=\frac{F(0)}{\left(1-\frac{q^2}{M_{\Sigma_b}^2}\right)
  \left[1-a\left(\frac{q^2}{M_{\Sigma_b}^2}\right)
  +b\left(\frac{q^2}{M_{\Sigma_b}^2}\right)^2\right]},
 \end{eqnarray}
where $F(q^2)$ stands for the form factors $f_{1,2}$ and $g_{1,2}$.
$a,~b$ and $F(0)$ in $F(q^2)$ are parameters which need to be fixed
using the form factors in the space-like region we calculate
numerically. This form can be automatically extended into the
time-like, i.e. physical region with $q^2\geq 0$.  The fitted values
of $a,~b$ and $F(0)$ in the form factors $f_{1,2}$ and $g_{1,2}$ are
presented in Table \ref{Tab:t2}. The dependence of the form factors
on $q^2$ is depicted in Fig. \ref{f2}.
\begin{table}
\caption{The $\Sigma_b\to \Sigma_c$ form factors given in the
  three-parameter form.}\label{Tab:t2}
\begin{ruledtabular}
\begin{tabular}{cccc}
  $F$    &  $F(0)$ &  $a$  &  $b$ \\
  $f_1$  &   0.4664     &   2.32    & 3.40   \\
  $f_2$  &     0.7358   &   2.08    &  2.08  \\
  $g_1$  &      -0.1298    &     1.15  &  0.42  \\
  $g_2$  &      -0.08977   &     1.11  & 1.07
\end{tabular}
\end{ruledtabular}
\end{table}

The values shown in Table \ref{Tab:t2} and Fig. \ref{f2} indicate
that the form factor $g_1$ and $g_2$ are small compared with $f_1$
and $f_2$ and  $f_1 (f_2)$ and $g_1 (g_2)$ have opposite signs, this
is similar to the case of $\Theta_b\rightarrow\Theta_c$
\cite{pentaquark2}.

Now let us turn to re-calculate the transition amplitude in the
HQET. In the heavy quark limit, we choose $\beta^\infty=0.50$ GeV
for $\Sigma_b$ and $\Sigma_c$.
%which is equal to the value for $\Sigma_b$.
The Isgur-Wise function is  parameterized as
 \begin{eqnarray}
 \xi(\omega)=1-\rho^2(\omega-1)+\frac{\sigma^2}{2}(\omega-1)^2+...,
 \end{eqnarray}
where $\rho^2\equiv-\frac{d\zeta(\omega)}{d\omega}|_{\omega=1}$ is
the slope parameter and
$\sigma^2\equiv\frac{d^2\zeta(\omega)}{d\omega^2}|_{\omega=1}$ is
the curvature of the Isgur-Wise function. Our fitted values are
 \beq
\xi_1=1-2.09(\omega-1)+1.84(\omega-1)^2\\
\xi_2=0.42[1-2.79(\omega-1)+3.09(\omega-1)^2].
 \eeq

%It is worth pointing out clearly that under the heavy quark limit
%where $1/M_Q$ correction is ignored, the relations among the form
%factors $f_i,\; g_i$ hold and thus one can easily determine the two
%Isgur-Wise functions from the six form factors which are calculated
%in terms of models (in this work we employ the LFQM). Our strategy
%here is that we still retain the relations, i.e. there are only two
%independent functions $\xi_1$ and $\xi_2$, but when we estimate
%them, we keep the mass of the light quarks to be finite and use the
%the experimentally fixed values for the masses of the heavy quarks.
%In this scheme, we in fact consider the $1/M_Q$ corrections.

\begin{figure}[hhh]
\begin{center}
\scalebox{0.8}{\includegraphics{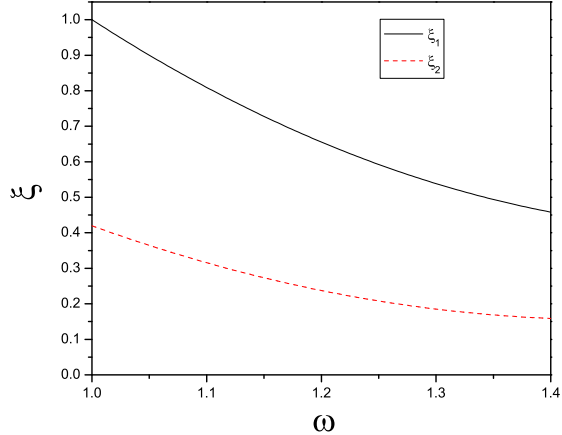}}
\end{center}
\caption{The $\Sigma_{b} \rightarrow \Sigma_{c}$ Isgur-Wise
function $\zeta(\omega)$ with diquark mass $m_{{[ud]}_V}=770$ MeV.
}\label{f3}
\end{figure}

The Isgur-Wise functions in the whole $\omega$ range is depicted in
Fig. \ref{f3}. One can notice that $\xi_1(\omega=1)=1$ holds as
required by the normalization of the Isgur-Wise function. Even
though, as indicated in literature, $\xi_2(\omega=1)$ is unknown, at
the large $N_C$ limit it is determined to be 1/2\cite{Chow:1994ni}
and other early model-dependent studies also confirm this
prediction\cite{Cheng:1996cs}.

It is worth indicating clearly that under the heavy quark limit,
i.e. $M_Q\to\infty$, the mass of heavy quark disappears in the
wavefunction (20), but the light constituent mass (anti-quark for
meson case and diquark for baryon case) remains. Therefore the
theoretical evaluation on the transition rate weakly depends on the
light constituent mass even under the heavy quark limit.

From Fig. \ref{f3}, we observe that $\xi_2|_{\omega=1}=0.42$ which
is slightly lower than $1/2$. This deviation is due to the mass
$m_{diquark}$ in the assumed wavefunction (see Eq.(21)). To further
explore the dependence, we deliberately vary $m_{{[ud]}_V}$,
$\beta_{b[ud]}$ and $\beta_{c[ud]}$. We find that $\xi_1$ does not
change at all, but the intercept $\xi_2(\omega=1)$ changes  for
different values of $m_{{[ud]}_V}$, $\beta_{b[ud]}$ and
$\beta_{c[ud]}$. For example, as $\xi_1=1$ and $\xi_2=0.47$ when one
sets $m_{{[ud]}_V}=0.5$ GeV, $\beta_{b[ud]}=0.4$ and
$\beta_{c[ud]}=0.35$. Definitely non-zero $m_{diquark}$ breaks the
heavy quark symmetry $SU_f(2)\otimes SU_s(2)$, but the violation is
still rather small, so that one can use the simplified expression
with only two Isgur-Wise functions to approach the transition matrix
elements.
% The errors in the parameter $\beta^\infty$ has only
%a minor effect which is consistent with the B meson case
%\cite{LWW}.
%In this work, we take the measured value as the input
%for the slope.

\subsection{Semi-leptonic decay of $\Sigma_b \to
\Sigma_c +l\bar{\nu}_l$}

%\begin{figure}[hhh]
%\begin{center}
%\scalebox{0.8}{\includegraphics{dv.eps}}
%\end{center}
%\caption{Differential decay width for $\Sigma_{b} \rightarrow
%\Sigma_{c}e\nu_e$ at $m_{{[ud]}_V}=770$ MeV. }\label{f4}
%\end{figure}
With the form factors given in last subsection, we are able to
calculate the width of $\Sigma_b \to \Sigma_c l\bar{\nu}_l$.
 %In
%Fig.\ref{f4} we present the differential decay width for
%$\Sigma_{b} \rightarrow \Sigma_{c}e\nu_e$.

In table \ref{Tab:t4} we list our numerical results. The predictions
are presented for two cases: with and without taking the heavy quark
limit.

It is also interesting to study the longitudinal and transverse
helicity amplitudes $ H^{V,A}_{\lambda',\lambda_W}$ where $\lambda'$
and $\lambda_W$ are the helicities of the daughter baryon and the
emitted W-boson respectively, since it may provide more information
about the model and even the whole framework. Moreover, several
asymmetry parameters $a_L$, $a_T$ and $P_L$ are defined in earlier
literature and in this work for readers' convenience we explicitly
present them in the appendix. A ratio of longitudinal to transverse
rates $R$ is also defined (see the appendix too), and $R>1$ implies
that the longitudinal polarization dominates. Because the values of
such asymmetries are more sensitive to the details of the employed
models, comparing the theoretical predictions on them with the data
which will be available soon at LHC as expected, can help to gain a
better understanding of the models.

In Tab.\ref{Tab:t4}  the predictions achieved with other approaches
\cite{Ebert:2006rp} are also presented. One notices from
Tab.\ref{Tab:t4} that there is an obvious discrepancy between
predictions on the semi-leptonic decay widths and $R$ values
estimated by different models. The future experimental measurements
would provide a chance to test the applicability of different
approaches.

\begin{table}
\caption{The widths (in unit $10^{10}{\rm s}^{-1}$) and polarization
asymmetries  of $\Sigma_b\to \Sigma_c l\bar{\nu}_l$ .}\label{Tab:t4}
\begin{ruledtabular}
\begin{tabular}{c|ccccccc}
    &  ${\rm width}$ &  $a_L$  &  $a_T$ &$\Gamma_L$& $\Gamma_T$ & $R$    &  $P_L$ \\\hline
 this work\footnote{ without the heavy quark limit}
   &    $1.38$      &   0.715 & -0.893&1.06& 0.32& 3.25 & 0.337 \\\hline
  this work\footnote{with the heavy quark limit}
    & $1.60$      &   0.706 & -0.966 &1.09&0.51& 2.13 & 0.171 \\\hline
 spectator-quark model$^a$ \cite{Singleton:1990ye}&   $4.3$       &   - & - &3.93 &0.37&10.7  & -\\\hline
  relativistic quark model$^b$\cite{Ebert:2006rp} &  $1.44$        &  -  & - &1.23 &0.21&5.89  & - \\\hline
the Bethe-Salpeter approach$^b$\cite{Ivanov:1998ya}&   $1.65$ &
- & - &- &-&- & -\\\hline
  relativistic three-quark model$^b$\cite{Ivanov:1996fj}&   $2.23$       &   - & - &1.90&0.33&5.76  & -
\end{tabular}
\end{ruledtabular}
\end{table}

%\begin{table}
%\caption{The widths (in unit $10^{-14}{\rm GeV}$) and polarization
%assymetries of $\Sigma_b\to \Sigma_c l\bar{\nu}_l$
%.}\label{Tab:t4}
%\begin{ruledtabular}
%\begin{tabular}{c|ccccccc}
%    &  $width$ &  $a_L$  &  $a_T$ &$\Gamma_L$& $\Gamma_T$ & $R$    &  $P_L$ \\\hline
 % within the heavy quark limit (this work)
 %   &    $0.91$      &   0.715 & -0.893&0.696& 0.214& 3.25 & 0.337 \\\hline
 % without the heavy quark limit (this work)
 %   & $1.05$      &   0.706 & -0.966 &0.714&0.336& 2.13 & 0.171 \\\hline
 % within the heavy quark limit (in \cite{EFG})
 %   &  $0.95$        &  -  & - &0.811 &0.139&5.87  & - \\\hline
  %  &   $1.08$       &   - & - &- &-&-  & -\\\hline
  %  &   $2.83$       &   - & - &- &-&-  & -\\\hline
 %   &   $1.47$       &   - & - & -&-&-  & -
%\end{tabular}
%\end{ruledtabular}
%\end{table}

\subsection{Non-leptonic decays of $\Sigma_b\to\Sigma_c+ M$}
From the theoretical aspects, calculating the concerned quantities
of the non-leptonic decays seem to be much more complicated than the
semi-leptonic ones. Our theoretical framework is based on the
factorization assumption, namely the hadronic transition matrix
element is factorized into a product of two independent matrix
elements of currents. One of them is determined by a decay constant
whereas the other is decomposed into a sum of a few terms according
to the Lorentz structure of the current and their coefficients are
the to-be-determined form factors.
%For the weak decays of mesons, such factorization
%approach seems to work well. We would believe that the factorization
%approach also works for the baryon case, especially as the diquark
%picture is employed.
The decays $\Sigma_b^0\to\Sigma_c+M^-$ is the so-called
color-favored transition, thus and factorization should be a good
approximation. Therefore, the study on these non-leptonic decays can
be a check of the consistency of the obtained form factors in the
heavy bottomed baryon system.

The formulas of the decay rates for non-leptonic decays
$\Sigma_b\to\Sigma_c +M$ in the factorization approach are given
in Ref.\cite{KK} and collected in our previous paper
\cite{Ke:2007tg} . Our numerical results are shown in
Tab.\ref{Tab:t6}. The CKM matrix elements, the effective Wilson
coefficient $a_1= 1$ and the meson decay constants are the same as
in Ref.\cite{Ke:2007tg}.

\ref{Tab:t6}.
%The Tables \ref{Tab:t7} and
%\ref{Tab:t8} demonstrate comparisons of our result with that in
%other approaches.
Two comments are made:

(1) The ratio $\frac{BR(\Sigma_b^0\to\Sigma_c^+
l^-\bar\nu_l)}{BR(\Sigma_b^0 \to\Sigma_c^+\pi^-)}$ is $11.4$ which
will be experimentally tested.

(2)The up-down asymmetry $\alpha$ for $\Sigma_b\to\Sigma_c V$ is
negative but that for $\Sigma_b\to\Sigma_c P$ is positive where
$\alpha$ is defined in the appendix .

\begin{table}
\caption{widths (in unit $10^{10}{\rm s}^{-1}$) and up-down
asymmetries of non-leptonic decays $\Sigma_b\to\Sigma_c M$ with
the light diquark mass $m_{[ud]}=500$ MeV.}\label{Tab:t6}
\begin{ruledtabular}
\begin{tabular}{c|cc|cc|}
  & \multicolumn{2}{c|}{without the heavy quark limit~~~~~~}
  & \multicolumn{2}{c|}{with the heavy quark limit~~~~} \\\hline
  & $\rm width$                & $\alpha$ & $\rm width$ & $\alpha$          \\\hline
  $\Sigma_b^0\to\Sigma_c^+ \pi^-$  & $0.140$  & $0.514$
                        & $0.146$  & $0.556$      \\\hline
  $\Sigma_b^0\to\Sigma_c^+ \rho^-$ & $0.166$  & $-0.653$
                        & $0.0907$  & $-0.785$  \\\hline
  $\Sigma_b^0\to\Sigma_c^+ K^-$    & $0.0115$  & $0.510$
                        & $0.0118$  & $0.548$      \\\hline
  $\Sigma_b^0\to\Sigma_c^+ K^{*-}$  & $0.00864$  & $-0.629$
                        & $0.00471$  & $-0.750$  \\\hline
  $\Sigma_b^0\to\Sigma_c^+ a_1^-$  & $0.163$  & $-0.551$
                        & $0.0880$  & $-0.646$  \\\hline
  $\Sigma_b^0\to\Sigma_c^+ D_s^-$  & $0.796$ & $0.379$
                        & $0.655$ & $0.425$  \\\hline
  $\Sigma_b^0\to\Sigma_c^+ D^{*-}_s$& $0.292$ & $-0.302$
                        & $0.152$  & $-0.317$  \\\hline
  $\Sigma_b^0\to\Sigma_c^+ D^-$  & $0.0266$ & $0.408$
                        & $0.0242$  & $0.440$  \\\hline
  $\Sigma_b^0\to\Sigma_c^+ {D^*}^-$& $0.0137$& $-0.331$
                        & $0.00694$  & $-0.356$
\end{tabular}
\end{ruledtabular}
\end{table}

\subsection{Estimate on the transition of $\Omega_b\to \Omega_c$}
Though we focus on the transition of $\Sigma_b\to\Sigma_c$ in this
work, the formulas deduced in section II can be applied to calculate
the transition between the $\frac{1}{2}^+$ baryons such as
$\Omega_b$ and $ \Omega_c$ whose structure is analogous to
$\Sigma_{b(c)}$ in the quark-diquark picture i.e. the diquark is an
axial vector.
% Under the quark-diquark picture the structure of
%$\Omega_{b(c)}$  i.e. the diquark [ss] in $\Omega_{b(c)}$ is an
%axial vector and its $J^P$ is $\frac{1}{2}^+$ so the formula
%deduced for the transition $\Sigma_b\to\Sigma_c$ can be employed
%to the transition $\Omega_b\to \Omega_c$.
Since the light diquark is regarded as a spectator, under the SU(3)
symmetry of light quarks the predictions on the  decay rates of
$\Sigma_b\to\Sigma_c$  hold for  $\Omega_b\to \Omega_c$
approximatively. As $\Omega_b$ decays via only weak interaction the
branching ratios of $\Omega_b\to \Omega_c$ should be dominant, so
that these decays can be detected more easily. Undoubtedly, since
the SU(3) symmetry is slightly broken,  different input parameters
would bring up minor differences for the numerical results of
$\Omega_b\to \Omega_c$ from $\Sigma_b\to\Sigma_c$, but the deviation
should be relatively small, and the allegation is supported  by some
theoretical studies which compare the width of $\Omega_b\to \Omega_c
e\nu$ with $\Sigma_b\to\Sigma_c e\nu$ (Tab. \ref{Tab:t7}).
\begin{table}
\caption{Various theoretical predictions on the rates $\Omega_b\to
\Omega_c e\nu$ and $\Sigma_b\to\Sigma_c e\nu$ (in unit $10^{10}{\rm
s}^{-1}$)}\label{Tab:t7}
\begin{ruledtabular}
\begin{tabular}{ccccccc}
  Decay & \cite{Ebert:2006rp} & \cite{Singleton:1990ye} & \cite{Ivanov:1996fj} & \cite{Ivanov:1998ya}\\\hline
  $\Sigma_b\to\Sigma_c e\nu$  & 1.44 &4.3  & 2.23    & 1.65   \\\hline
 $\Omega_b\to
\Omega_c e\nu$    & 1.29  & 5.4   & 1.87 & 1.81
\end{tabular}
\end{ruledtabular}
\end{table}

\section{Conclusions and discussions}

In this paper, we extensively explore the $\Sigma_b\to\Sigma_c$
transition  in all details  and estimate  the widths for the
semi-leptonic decay and non-leptonic two-body decays of
$\Sigma_b\to\Sigma_c $ as well as several relevant measurable
quantities. For the heavy baryons the quark-diquark picture is
employed, which reduces the three-body structure into a two-body
one.

The matrix elements of  the transition $\Sigma_b\to\Sigma_c$ can be
parameterized with a few form factors $f_i$ and $g_i$ ($i=1,2,3$)
according to the Lorentz structures, and we obtain these form
factors by calculating the transition $\Sigma_b\to\Sigma_c$ in the
LFQM and evaluate them numerically. The form factors $f_1$ and $f_2$
for $\Sigma_b\to\Sigma_c$ are much larger than $g_1$ and $g_2$, it
is  noted that $f_1 (f_2)$ and $g_1 (g_2)$ have opposite signs.
Furthermore, we also derive the generalized Isgur-Wise functions
$\xi_1$ and $\xi_2$ under the heavy quark limit. We find that
$\xi_1(\omega=1)=1$ is consistent with the normalization condition,
but $\xi_2(\omega=1)$ is slightly lower than 1/2 which was predicted
by large $N_c$ theory. Our analysis indicates that the deviation is
due to the non-zero mass of the light constituents in hadrons (meson
and baryon). With the form factors derived in terms of the LFQM or
the Isgur-Wise functions we evaluate the semi-leptonic decay rates
of $\Sigma_b\to\Sigma_c$ with and without taking the heavy quark
limit. The results with and without heavy quark limit do not decline
much from each other, moreover, our numerical results of the rates
are generally consistent with that estimated by different
approaches. However, it is interesting to note that for the
transverse polarization asymmetry $P_L$, there is an obvious
discrepancy between our results and those by other approaches.
Moreover, in terms of SU(3) symmetry of light quarks we estimate the
rates $\Omega_b\to\Omega_c$ which is approximately equal to those of
$\Sigma_b\to\Sigma_c$.

Since the LHCb is running successfully and a remarkable amount of
data on $\Sigma_b$ and $\Omega_b$ production and decay is being
accumulated, especially by the LHCb detector, thus we have all
confidence that in near future (maybe not next year, but anyhow
won't be too far away), their decay rates and even the asymmetries
would be more accurately measured, and we will have a great
opportunity to testify our models.

Now let us estimate the feasibility of observing the decay process
$\Sigma_b\rightarrow\Sigma_cl\nu_l$. Firstly, we use the code
PYTHIA8.1 to calculate the production  cross section of $\Sigma_b^+$
via  $pp\rightarrow b\bar b\rightarrow\Sigma_b^++X$. By the
PYTHIA8.1\cite{Bargiotti:2007zz,Brambilla:2010cs,Sjostrand:2007gs},
100000 $b\bar b$ pairs are generated at the
$E_\mathrm{CM}=7\mathrm{TeV}$. Then  $N_{\Sigma_b^+}=354$ are
produced and the corresponding production cross section is
$\sigma_{\Sigma_b^+}\approx354/100000=3.54\times10^{-3}\sigma_{b\bar
b}$. In 2011, the integrated luminosity of the LHCb is
$1\;fb^{-1}$\cite{Rodrigues:2011zz} and the production cross section
of the $b\bar b$ pairs is $\sigma_{b\bar b}=288\mu
b$\cite{Aaij:2011jh}, so our estimate is that  about
$1.02\times10^9$ $\Sigma_b^+$ exist in the 2011 data.

Because the LHCb detector is good at tracking the charged particles,
such as $p^+$, $K^\pm$ and $\pi^\pm$ and charged leptons, we suggest
the decay chain used to find the $\Sigma_b$'s semileptonic decay is
that: $\Sigma_b^+\rightarrow\Sigma_c^0\mu^+\nu_\mu$,
$\Sigma_c^0\rightarrow\Lambda_c^+\pi^-$ (the branching ratio is
$\Gamma_{\Lambda_c^+\pi^-}/\Gamma_{\Sigma_c}\approx100\%$) and
$\Lambda_c^+\rightarrow pK^+\pi^-$ (the branching ratio is
$\Gamma_{pK^+\pi^-}/\Gamma_{\Lambda_c}\approx5\%$). So, the
measurable number of this decay chain is:
\begin{eqnarray}\label{num of sigma_b}
\mathcal{N}_{\Sigma_b}=1.02\times10^9\times{\Gamma_{\Sigma_c\mu\nu_\mu}
\over
\Gamma_{\Sigma_b}}\times100\%\times5\%\times\epsilon_\mathrm{trig}\times\epsilon_\theta,
\end{eqnarray}
where, ${\Gamma_{\Sigma_c\mu\nu_\mu}/\Gamma_{\Sigma_b}}$ stands for
the branching ratio of $\Sigma_b$'s semi-leptonic decay, the
$\epsilon_\mathrm{trig}$ is the efficiency of the detection trigger
and $\epsilon_\theta\approx20\%$ is the efficiency of the detector's
geometric acceptance\cite{He:2010zqa}. Without losing generality, we
set $\epsilon_\mathrm{trig}=88\%$ here (also given in
\cite{He:2010zqa}). The decay width of $\Sigma_b$ is not known yet.
Since QCD is flavor-blind, thus we have reason to believe that
considering the phase space of final state the decay width of the
$\Sigma_b$ should be related to that of $\Gamma_{\Lambda_b}$,
$\Gamma_{\Lambda_c}$ and $\Gamma_{\Sigma_c}$ as:
\begin{eqnarray}\label{decay width}
 \Gamma_{\Sigma_b}=\Gamma_{\Lambda_b}{\Gamma_{\Sigma_c}\over\Gamma_{\Lambda_c}}=
4.81\times10^{10}({\,\rm in \,unit} 10^{10}s^{-1}). %0.3MeV
\end{eqnarray}
Substituting  all these values back into Eq.(\ref{num of sigma_b})
( $\Gamma_{\Sigma_c\mu\nu_\mu}=1.38\times10^{10}s^{-1}$ can be
found in Tab-\ref{Tab:t4}), we have the number of signals of
$\Sigma_b\to\Sigma_c+l\bar\nu$:
\begin{eqnarray}
 \mathcal{N}_{\Sigma_b}=1.87\times10^{-4}\frac{\Gamma_{\Sigma_c\mu\nu_\mu}}{10^{10}s^{-1}}=2.58\times10^{-4}.
\end{eqnarray}
If the luminosity of LHCb is not increased greatly in the future to
avoid the high level pile-up, we conclude that, since the strong
decay $\Sigma_b\rightarrow\Lambda_b\pi$  dominates and the lifetime
of $\Sigma_b$  is determined by the mode, the branching ratio of the
weak decay is significantly suppressed, it would be hard to directly
observe the signals of semileptonic decays of $\Sigma_b$. Since the
signal of the semileptonic decays of $\Sigma_b$ is clear and related
to new physics,  so that is worth careful investigation at LHCb,
even though it is almost impossible to be measured for the present
luminosity if only SM applies. Thus, as analyzed above, we would
recommend to measure $\Omega_b\to\Omega_c$ transitions because
$\Omega_b$ does not decay via strong interaction, so
$\Omega_b\to\Omega_c+l\bar\nu$ and $\Omega_b\to\Omega_c+M$ would be
the dominant modes. It enables us to make a more precise measurement
by which we can  not only further investigate the validity of the
diquark picture for heavy baryons, but also create an opportunity to
search for new physics beyond the SM, at least check if the new
physics scenario shows up in such transitions. We also suggest to
measure the quantities such as the asymmetries besides the widths,
because of the obvious advantages about our models and physics.

\section*{Acknowledgement}

This work is supported by the National Natural Science Foundation
of China (NNSFC) under the contract No. 11075079, No. 11175091 and
No. 11005079; the Special Grant for the Ph.D. program of Ministry
of Eduction of P.R. China No. 20100032120065.

\appendix

\section{Semi-leptonic decays of
 $\Sigma_b \to \Sigma_c l\bar\nu_l$ }

The helicity amplitudes are related to the form factors for
$\Sigma_b\to\Sigma_c$ through the following expressions \cite{KKP}
 \beq
 H^V_{\frac{1}{2},0}&=&\frac{\sqrt{Q_-}}{\sqrt{q^2}}\left(
  \left(\Mb+\Mc\right)f_1-\frac{q^2}{\Mb}f_2\right),\non\\
 H^V_{\frac{1}{2},1}&=&\sqrt{2Q_-}\left(-f_1+
  \frac{\Mb+\Mc}{\Mb}f_2\right),\non\\
 H^A_{\frac{1}{2},0}&=&\frac{\sqrt{Q_+}}{\sqrt{q^2}}\left(
  \left(\Mb-\Mc\right)g_1+\frac{q^2}{\Mb}g_2\right),\non\\
 H^A_{\frac{1}{2},1}&=&\sqrt{2Q_+}\left(-g_1-
  \frac{\Mb-\Mc}{\Mb}g_2\right).
 \eeq
where $Q_{\pm}=2(P\cdot P'\pm \Mb\Mc)$. The amplitudes for the
negative helicities are obtained in terms of the relation
 \beq
 H^{V,A}_{-\lambda'-\lambda_W}=\pm H^{V,A}_{\lambda',\lambda_W},
  \eeq
where the upper (lower) sign corresponds to V(A).
 The helicity
amplitudes are
 \beq
 H_{\lambda',\lambda_W}=H^V_{\lambda',\lambda_W}-
  H^A_{\lambda',\lambda_W}.
 \eeq
The helicities of the $W$-boson $\lambda_W$ can be either $0$ or
$1$, which correspond to the longitudinal and transverse
polarizations, respectively.  The longitudinally (L) and
transversely (T) polarized rates are respectively\cite{KKP}
 \beq
 \frac{d\Gamma_L}{d\omega}&=&\frac{G_F^2|V_{cb}|^2}{(2\pi)^3}~
  \frac{q^2~p_c~\Mc}{12\Mb}\left[
  |H_{\frac{1}{2},0}|^2+|H_{-\frac{1}{2},0}|^2\right],\non\\
 \frac{d\Gamma_T}{d\omega}&=&\frac{G_F^2|V_{cb}|^2}{(2\pi)^3}~
  \frac{q^2~p_c~\Mc}{12\Mb}\left[
  |H_{\frac{1}{2},1}|^2+|H_{-\frac{1}{2},-1}|^2\right].
 \eeq
where $p_c$ is the momentum of $\Sigma_c$ in the reset frame of
$\Sigma_b$.

The integrated longitudinal and transverse asymmetries defined as
 \beq
 a_L&=&\frac{\int_1^{\omega_{\rm max}} d\omega ~q^2~ p_c
     \left[ |H_{\frac{1}{2},0}|^2-|H_{-\frac{1}{2},0}|^2\right]}
     {\int_1^{\omega_{\rm max}} d\omega ~q^2~ p_c
     \left[|H_{\frac{1}{2},0}|^2+|H_{-\frac{1}{2},0}|^2\right]},
     \non\\
 a_T&=&\frac{\int_1^{\omega_{\rm max}} d\omega ~q^2~ p_c
     \left[ |H_{\frac{1}{2},1}|^2-|H_{-\frac{1}{2},-1}|^2\right]}
     {\int_1^{\omega_{\rm max}} d\omega ~q^2~ p_c
     \left[|H_{\frac{1}{2},1}|^2+|H_{-\frac{1}{2},-1}|^2\right]}.
 \eeq
 The ratio of the longitudinal to
transverse decay rates $R$ is defined by
 \beq
 R=\frac{\Gamma_L}{\Gamma_T}=\frac{\int_1^{\omega_{\rm
     max}}d\omega~q^2~p_c\left[ |H_{\frac{1}{2},0}|^2+|H_{-\frac{1}{2},0}|^2
     \right]}{\int_1^{\omega_{\rm max}}d\omega~q^2~p_c
     \left[ |H_{\frac{1}{2},1}|^2+|H_{-\frac{1}{2},-1}|^2\right]},
 \eeq
and the  longitudinal $\Sigma_c$ polarization asymmetry $P_L$ is
given as
 \beq
 P_L&=&\frac{\int_1^{\omega_{\rm max}} d\omega ~q^2~ p_c
     \left[ |H_{\frac{1}{2},0}|^2-|H_{-\frac{1}{2},0}|^2+
     |H_{\frac{1}{2},1}|^2-|H_{-\frac{1}{2},-1}|^2\right]}
     {\int_1^{\omega_{\rm max}} d\omega ~q^2~ p_c
     \left[|H_{\frac{1}{2},0}|^2+|H_{-\frac{1}{2},0}|^2+
     |H_{\frac{1}{2},1}|^2+|H_{-\frac{1}{2},-1}|^2\right]}.
 \eeq

\end{document}